\documentclass[a4paper,UKenglish,cleveref, autoref, thm-restate]{lipics-v2021}
\hideLIPIcs  %uncomment to remove references to LIPIcs series (logo, DOI, ...), e.g. when preparing a pre-final version to be uploaded to arXiv or another public repository

\title{Time Moves Faster When There is Nothing You Anticipate: The Role of Time in MEV Rewards} %TODO Please add

\titlerunning{The Role of Time in MEV Rewards} %TODO optional, please use if title is longer than one line

\usepackage{pdfpages}

\author{Burak {\"Oz}}{Technical University of Munich, Boltzmannstrasse 3, 85748 Garching, Germany}{burak.oez@tum.de}{}{}
\author{Benjamin {Kraner}}{University of Zurich, Andreasstrasse 15, 8050 Zurich, Switzerland}{benjamin.kraner@uzh.ch}{}{}
\author{Nicolò {Vallarano}}{University of Zurich, Andreasstrasse 15, 8050 Zurich, Switzerland}{nicolo.vallarano@uzh.ch}{}{}
\author{Bingle Stegmann {Kruger}}{University of Cape Town, University Avenue 18, 7700 Rondebosch, Cape Town, South Africa}{krgbin001@myuct.ac.za}{}{}
\author{Florian {Matthes}}{Technical University of Munich, Boltzmannstrasse 3, 85748 Garching, Germany}{florian.matthes@tum.de}{}{}
\author{Claudio Juan {Tessone}}{University of Zurich, Andreasstrasse 15, 8050 Zurich, Switzerland}{claudio.tessone@uzh.ch}{}{}

\authorrunning{B. Öz, B. Kraner, N. Vallarano, B.S. Kruger, F. Matthes, and C.J.Tessone} %TODO mandatory. First: Use abbreviated first/middle names. Second (only in severe cases): Use first author plus 'et al.'

\Copyright{Jane Open Access and Joan R. Public} %TODO mandatory, please use full first names. LIPIcs license is "CC-BY";  http://creativecommons.org/licenses/by/3.0/


 %TODO mandatory; please add comma-separated list of keywords

\category{} %optional, e.g. invited paper

\relatedversion{} %optional, e.g. full version hosted on arXiv, HAL, or other respository/website
\nolinenumbers %uncomment to disable line numbering

\EventEditors{John Q. Open and Joan R. Access}
\EventNoEds{2}
\EventLongTitle{Advances in Financial Technologies (AFT’23)}
\EventShortTitle{AFT’23}
\EventAcronym{AFT’23}
\EventYear{2023}
\EventDate{December 24--27, 2016}
\EventLocation{Little Whinging, United Kingdom}
\EventLogo{}
\SeriesVolume{42}
\ArticleNo{}
\begin{document}

\maketitle

\begin{abstract}

This study explores the intricacies of waiting games, a novel dynamic that emerged with Ethereum's transition to a Proof-of-Stake (PoS)-based block proposer selection protocol. Within this PoS framework, validators acquire a distinct monopoly position during their assigned slots, given that block proposal rights are set deterministically, contrasting with Proof-of-Work (PoW) protocols. Consequently, validators have the power to delay block proposals, stepping outside the honest validator specs, optimizing potential returns through MEV payments. Nonetheless, this strategic behaviour introduces the risk of orphaning if attestors fail to observe and vote on the block timely. Our quantitative analysis of this waiting phenomenon and its associated risks reveals an opportunity for enhanced MEV extraction, exceeding standard protocol rewards, and providing sufficient incentives for validators to play the game. Notably, our findings indicate that delayed proposals do not always result in orphaning and orphaned blocks are not consistently proposed later than non-orphaned ones. To further examine consensus stability under varying network conditions, we adopt an agent-based simulation model tailored for PoS-Ethereum, illustrating that consensus disruption will not be observed unless significant delay strategies are adopted. Ultimately, this research offers valuable insights into the advent of waiting games on Ethereum, providing a comprehensive understanding of trade-offs and potential profits for validators within the blockchain ecosystem.

\end{abstract}

\section{Introduction}
\label{sec:introduction}
The blockchain landscape has seen tremendous growth and evolution over the years, with innovative concepts continually emerging to enhance its functionality and efficiency. Ethereum, a key player in this arena, has recently undergone major upgrades, with the merge\footnote{\url{https://ethereum.org/en/roadmap/merge/}} being the most prevalent, replacing the Proof-of-Work (PoW)-based block proposer selection mechanism with a Proof-of-Stake (PoS)-based one~\cite{spychiger2021unveiling, Tasca_Tessone_2019}. With this upgrade, Etheruem also introduced a new consensus protocol Gasper comprising Latest Message Driven Greediest Heaviest Observed SubTree (LMD GHOST) for fork-choice and Casper Friendly Finality Gadget (FFG) for finality, replacing the longest-chain consensus. Switching to PoS remarkably reduced the electricity consumption of the Ethereum blockchain (down by 99.98\% according to the 2022 report of Crypto Carbon Ratings Institute\footnote{\url{https://indices.carbon-ratings.com/ethereum-merge}}) which led to a significant cut down in block proposal rewards as consensus participants, known as validators, do not require as high incentives as PoW-Ethereum miners did. Hence, the validators depend even more on transaction fees and exogenous rewards like Maximal Extractable Value (MEV), then miners did.

Incentives play an essential role in the evolution of public, permissionless blockchains such as Ethereum. Misaligned incentives, however, can lead to a lack of interest in contributing to a network or, potentially, to profit-seeking attacks, dangering consensus stability \cite{carlsten}. MEV has emerged as a powerful incentive within Ethereum~\cite{fb}, enabling network participants to gain profits beyond protocol rewards through strategic transaction issuance and ordering~\cite{oz_advent_2023}. In particular, participants who partake in MEV extraction employ transaction ordering techniques inspired by traditional finance to execute profitable strategies based on the network state concerning pending transactions in the mempool and the most recently finalized blockchain state~\cite{qin_quant}. The prominence of MEV has risen with the advent of Decentralised Finance (DeFi) ecosystem, as shown by the emergence of sophisticated bots exploiting MEV opportunities, regardless of their potential disruption to users \cite{fb}.

% For example, Maximal Extractable Value (MEV)\footnote{Previously the term was called Miner Extractable Value, coined by Daian et al.~\cite{fb}.} remains a powerful incentive within the Ethereum blockchain, which allows any network participant to extract additional value (i.e.~profits) beyond the standard protocol rewards by influencing transaction ordering through payments to block proposers~\cite{oz_advent_2023}. Notwithstanding, while MEV always existed in public, permissionless blockchains, it has become more prominent since 2020 with the emergence of DeFi. Articles such as Ethereum is a Dark Forest~\cite{Robinson2020} and Escaping the Dark Forest~\cite{Samczsun2020} introduced MEV to a public audience and demonstrated how dangerous the MEV game can be. Specifically, the danger lies in the fact that there are many highly complex bots that attempt to profit from any possible opportunity exposed to the network through submitted transactions without regard for the disruptive impact on users. 

Currently, the PoS-Ethereum protocol allows validators to release their block in a 12 seconds slot time, during which the designated validator has a total monopoly on the block release. This feature introduces new opportunities for MEV extraction through strategic \textit{waiting games}. To illustrate, we hypothesize that the block proposer may exploit their monopolistic position and wait as long as possible within the slot limits to release the block in order to extract additional MEV. However, if a large proportion of validators on the Ethereum network engage in these delayed releases, it could result in instability within the consensus of the network, as delays in block release may lead to multiple forked blocks. Thus, this paper delves into the impact of delayed block releases, or more broadly, waiting games, on the Ethereum consensus framework. We primarily focus on answering three pivotal research questions:
\begin{enumerate}
    \item Are waiting games profitable for validators, considering the dynamics of value evolution during a slot time and the relative importance of MEV rewards compared to proposal rewards from the consensus layer?
    \item Can validators engage in waiting games without risking their blocks being vulnerable to getting orphaned?
    \item Could waiting games potentially disrupt the stability of Ethereum consensus and, if so, under which conditions?
\end{enumerate}

\subsection{Related Work}
As far as we are aware, \cite{schwarzschilling2023time} is the sole study exclusively focusing on waiting or timing games in PoS-Ethereum, although \cite{wahrstatter_time_2023} also notes a positive correlation between bid arrival time and value, while exploring the intricate details about the block construction market available on Ethereum, MEV-Boost \cite{mevboost}. In \cite{schwarzschilling2023time}, the authors present a model where honest-but-rational consensus participants can delay their block proposals to maximise MEV capture, while still ensuring their inclusion in the blockchain in a timely manner. Despite noting that timing games are worthwhile, they find these strategies not currently being exploited by consensus participants. However, drawing attention to the differences between our study and that of the authors in \cite{schwarzschilling2023time} is critical. While we also analyze the value of time, we further investigate the importance of MEV rewards relative to consensus layer proposal rewards and examine unrealised values due to validators' acting prematurely, not capturing the maximum payments by the builders. Additionally, we assess attestations concerning forking vulnerability and explore the connection between orphaned blocks and the timing of their winning bids, determining if orphaned blocks arrive later than others or if there are non-orphaned blocks with later submitted winning bids. This comprehensive inspection provides deeper insights into the value and significance of waiting games and the associated risks.

To test the waiting games under different network conditions, we adopt an Agent-Based Model (ABM) specifically devised for PoS-Ethereum by~\cite{kraner2023agentbased}. In particular, studying the impact of waiting games on the Ethereum consensus requires a robust model to examine how different waiting strategies influence the consensus properties. Moreover, latency plays an essential role in defining consensus. Therefore, the ABM model from~\cite{kraner2023agentbased} is ideal as the model explicitly includes variables for the topology of the peer-to-peer network, and of particular interest, the average block and attestation latencies. Furthermore, the ABM provides specific techniques for accurately measuring the consensus of the simulated PoS-Ethereum network.

\subsection{Contribution}
This work delivers the following contributions:
\begin{enumerate}
    \item Illuminates the potential profitability and inherent risks of waiting games in PoS-Ethereum, offering new insights into block proposal strategies and validator trade-offs.
    \item Underlines the economic impact of MEV rewards, revealing their influence on rational proposer behaviour and showcasing unrealised value within premature bid selection.
    \item Delivers a comprehensive analysis of attestation shares, identifying blocks at higher forking risk, and contrasts orphaned and non-orphaned blocks, highlighting the impact of proposal timing on confirmation.
    \item Employs an agent-based model to simulate various validator behaviour scenarios in Ethereum consensus, enhancing our understanding of its resilience and stability against timing games.
\end{enumerate}

\subsection{Organization} Beyond this introduction, the paper is organized into four additional sections. Section~\ref{sec:background} presents essential background information on PoS-Ethereum mechanisms, MEV, and Proposer-Builder Separation (PBS). This is followed by Section~\ref{sec:methodology}, which details the methodology adopted for data collection and processing. The findings from our empirical analysis and agent-based simulation are presented in Section~\ref{sec:results}. We conclude the paper and summarize our findings in Section~\ref{sec:conclusion}.

\section{Background}
\label{sec:background}

% The purpose of this section is to present the pertinent background literature related to the work conducted in this study. Specifically, this section briefly describes the Proof-of-Stake (PoS) Ethereum consensus, followed by the reward structure of PoS-Ethereum. Subsequently, we introduce Maximal Extractable Value (MEV), which refers to the additional profit that can be gained through strategic transaction ordering, and is a key focus of our study. Furthermore, the next section presents Proposer-Builder Separation (PBS) - i.e., the separation of the roles of proposers (validators) and builders (who construct blocks) to counter MEV centralisation risks. 

This section aims to provide relevant background information essential to our study. Initially, we outline the Proof-of-Stake (PoS) Ethereum consensus and follow it up with a discussion on the reward structure within PoS-Ethereum. Next, we delve into the concept of Maximal Extractable Value (MEV), an additional profit that network participants can capture through certain strategies employed during block production, which forms a crucial aspect of our research. Lastly, we present the Proposer-Builder Separation (PBS)-the separation of the roles in block production to proposers (validators) and builders (who construct blocks)-to mitigate the risks associated with MEV centralization around the consensus layer.

\subsection{PoS-Ethereum Consensus}

The Ethereum network is a peer-to-peer network that operates in a decentralised manner and comprises nodes (computers or servers) that maintain a copy of the blockchain. Moreover, the nodes can offer varying functionalities, such as full nodes or light nodes. However, certain nodes perform a specific role as validators, incentivised by reward mechanisms. Proof-of-Stake (PoS) is a blockchain mechanism used to select the next block proposer in a Sybil attack-resistant manner. In PoS, network members who wish to participate in block production must stake money to the protocol, and unlike PoW, there is no need for participants to solve a complex puzzle or use mining hardware. Instead, participants have a financial stake that can be slashed in case of misbehaviour. This makes PoS chains more energy-efficient compared to PoW chains.

In the PoS-Ethereum consensus, validators are participants involved in the consensus process. Validators need to deposit a stake as collateral (32 Ether) to the deposit contract, which contains a Merkle root for all deposits. In addition, validators are required to maintain a node that facilitates communication and message exchange within the gossip network. Validators are incentivised by the rewards they receive for honest conduct, thereby compensating them for their associated costs. The protocol randomly selects a validator to propose a block. If a validator misbehaves (e.g.~proposes multiple blocks in one slot), the respective validator gets penalised (i.e.~stake slashed). Additionally, a committee of validators is chosen at every slot to vote on the validity of the proposed block. At least \textit{two third} of the committee must agree for the block to be considered valid. Validators who submit contradicting votes are penalised. The PoS-Ethereum protocol divides time into epochs and slots, with each epoch containing 32 slots and each slot lasting 12 seconds. In PoS-Ethereum, a block is considered finalised after two epochs. 

Validators in Ethereum use the LMD GHOST~\cite{Sompolinsky2015} fork choice rule to establish the canonical chain and accurately propose or attest to blocks. Under this framework, validators maintain a record of the latest message received from all other validators and only update it if the new vote comes from a slot strictly later than the existing entry. Therefore, in the event that a validator receives two equivocating votes from the same validator for the same slot, only the vote that arrives first is taken into consideration. The consensus mechanism operates in two phases. In the first phase, LMD GHOST operates on a smaller time scale, with time segments referred to as slots. In the second phase, Casper FFG~\cite{buterin2019casper} operates on a larger time scale, within epochs composed of 32 slots. Specifically, Casper FFG is a leaderless, two-phase propose-and-vote-style Byzantine fault-tolerant consensus protocol that functions to finalise blocks and ensure security even during temporary network partitions. The confirmation rule at the Casper FFG level involves outputting the most recent finalised block and its prefix.

\subsection{Reward structure}

% \textcolor{orange}{It would be interesting to analyze the following rewards:
% \begin{itemize}
%     \item Reward from Staking
%     \item Reward from Gas
%     There may be a relationship between the time delay of a block and the size of the Gas reward. How full are the blocks of ethereum? If we show, that gas reward is small this is not very important
%     \item Reward from MEV
%     Well we do that in the rest of the document
% \end{itemize}
% We do not need an empirical analysis...}

In the post-merge Ethereum, there have been considerable changes in the reward structure. This paper provides a brief overview of these changes while directing interested readers to~\cite{cortesgoicoechea2023autopsy} for more detailed information. Validators earn rewards by casting votes that align with the majority of the other validators (i.e.~honest interaction), specifically when proposing blocks and participating in syncing committees~\cite{ethconsensusspecs}. Moreover, validators receive rewards for attesting and proposing blocks on time (e.g.~attestation rewards). As a result, the network reaches a consensus and validates the data in the blockchain (i.e.~finalisation). Block proposers receive an aggregated reward of the \textit{Gas Tips} (i.e.~the fees users pay to prioritise their transactions) included in a block. On the other hand, the value of the attestation rewards for each epoch is calculated based on a $Base\_Reward$, which functions as the fundamental unit for the other rewards. The $Base\_Reward$ is determined by the validator's effective balance and the total number of active validators and represents the average reward that a validator can receive per epoch when operating under optimal conditions. Specifically, the calculation entails dividing the product of the validator's effective balance and a $Base\_Reward\_Factor$ by the square root of the total active balance, as shown in Equation~\ref{eq:basereward}:

\begin{equation}
\label{eq:basereward}
\text{Base\_Reward} = \frac{\text{Effective\_Balance} \times \text{Base\_Reward\_Factor}}{\sqrt{\text{Active\_Balance}}}
\end{equation}

Where the $\text{Effective}\_\text{Balance}$ is the effective balance of the attesting validator and  the $\text{Active\_Balance}$ is the total active balance staked across all active validators, and the $\text{Base\_Reward\_Factor}$ is a constant of value 64. Therefore, the $\text{Base\_Reward}$ is proportional to the effective balance of the validator and inversely proportional to the number of validators present. Accordingly, as the number of validators in the network increases, the rewards are distributed among more validators, which decreases the rewards per validator. Moreover, validators are rewarded based on the correctness of their behaviour. Specifically, the attestation reward is influenced by the Flag\_Reward~\cite{cortesgoicoechea2023autopsy}, shown in:

\begin{equation}
\label{eq:flagreward}
\text{Flag\_Reward} = \frac{\text{Flag\_Weight}}{\text{Weight\_Denom}} \times \text{Base\_Reward} \times \frac{\text{Attestation\_Balance}}{\text{Active\_Balance}}
\end{equation}

Where Flag\_Weight and Weight\_Denom are constants that vary between flags, and the Attestation\_Balance is the sum of the effective balances of all attesting validators. Accordingly, the Attestation\_Reward is the sum of the flag rewards ($\sum \text{Flag\_Reward}$). In addition, sync rewards and block proposal rewards form part of the reward structure. The former is a reward for validators that participate in the sync committee to sign new block headers in every slot. The latter are the rewards linked to proposing a beacon block. Specifically, the reward consists of including the attestation aggregations from validator votes not yet present in the beacon chain and including sync aggregates. Moreover, the block proposer is responsible for including the aggregate from the sync committee participants~\cite{cortesgoicoechea2023autopsy}.

% \subsection{Latency} 
% \textcolor{red}{TODO - is this required?}

% The technical infrastructure of Ethereum, as well as most permissionless blockchains, relies on a peer-to-peer (P2P) communication network because of the inherent flexibility and distribution possessed by P2P networks~\cite{neudecker2018network}. However, P2P networks sacrifice network performance, particularly in message delivery latency~\cite{decker2013information,delgado2018cryptocurrency}. 

% Nevertheless, the performance of existing permissionless blockchains is limited by the rate of block production and transaction confirmation, rather than network latency...consensus mechanism...

% For example, Ethereum 
% 12 seconds

%ETH p2p network
%How big is it (size)
%Shape
%Latency estimation - distribution
%Edge-to-edge

% \subsection{Agent-based Modeling}
% To test the waiting games under different network conditions, we adopt an Agent-Based Model (ABM) specifically devised for PoS-Ethereum by~\cite{kraner2023agentbased}. In particular, studying the impact of waiting games on the Ethereum consensus requires a robust model to examine how different waiting strategies influence the consensus properties. Moreover, latency plays an essential role in defining consensus. Therefore, the ABM model from~\cite{kraner2023agentbased} is ideal as the model explicitly includes variables for the topology of the peer-to-peer network, and of particular interest, the average block and attestation latencies. Furthermore, the ABM provides specific techniques for accurately measuring the consensus of the simulated PoS-Ethereum network.

\subsection{Maximal Extractable Value}
The concept of Maximal Extractable Value (MEV) has grown significantly within the Ethereum blockchain, particularly with the surge of the Decentralized Finance (DeFi) domain since 2020 \cite{oz_advent_2023}. Daian et al. \cite{fb} initially coined the term Miner Extractable Value, implying the additional value that miners, or privileged actors, could seize by manipulating transaction sets and order in a block, beyond standard protocol incentives such as block rewards and transaction fees \cite{fb}. Despite these privileged actors having a position of power, this value is also available to any network participant who can observe the confirmed blockchain state and monitor pending transactions. Consequently, the Miner or Privileged Extractable Value forms a subset of Maximal Extractable Value, which includes any value that can be extracted within a blockchain network \cite{qin_quant}.

To exploit MEV, network participants who do not possess authority over block content must influence the transaction order by offering payments to block proposers. These payments can be transaction fees or direct payments facilitated by MEV markets such as Flashbots Auction \cite{fbauction}. Participants seeking MEV, known as MEV searchers, employ transaction ordering techniques inspired by traditional finance, such as frontrunning and backrunning \cite{bracciali_sok_2020, frontrunnerjones, zhou_high-frequency_2021}, and DeFi instruments like flash loans \cite{qin_attacking_2021}, to execute profitable strategies based on the network state concerning pending transactions in the mempool \cite{zhou_high-frequency_2021} and the most recently confirmed blockchain state \cite{zhou_jit}.

Existing research \cite{qin_quant, fbPan} and web services such as MEV Explore \cite{mevexp} or EigenPhi \cite{eigen} that quantify the extracted MEV on Ethereum are limited by their heuristic approaches to detecting certain MEV patterns (e.g., arbitrages, sandwiches, liquidations) and the protocols they inspect. Consequently, they only provide a lower-bound estimate of the actual value extracted. According to MEX Explore \cite{mevexp}, more than \$675 million worth of MEV had been harnessed until the merge in September 2022. Furthermore, data from EigenPhi shows ongoing MEV extraction, with approximately \$9.5 million in profit generated within a month from May 2023 to June 2023 \cite{eigen}. Although a lower-bound, these estimates already highlight the significance of MEV as a powerful incentive on Ethereum, emphasizing its consideration in design decisions to prevent consensus destabilizing attacks \cite{carlsten, twoattacks, schwarz-schilling_three_2022}.

MEV's negative implications are manifold \cite{yang2022sok}, with issues including user value loss, network congestion due to MEV searchers competing \cite{fb}, value disparities among blocks incentivising consensus destabilising attacks \cite{carlsten}, and centralisation of MEV supply chain components due to economies of scale \cite{supply}. In response, innovative solutions are emerging across different system layers \cite{yang2022sok}. These include fair transaction ordering protocols \cite{noauthor_hedera_nodate, fair}, privacy-preserving mechanisms to protect valuable transaction information of users \cite{ noauthor_mev-share_nodate, fbsgx, Zhang2022F3BAL}, efficient MEV extraction designs that facilitate flow across different stakeholders \cite{flashbotsmev-boost_2023, suave}, and MEV-aware applications providing user rebates \cite{noauthor_cow_nodate, mevblocker, noauthor_mev-share_nodate}. For a comprehensive analysis of the MEV mitigation space and existing solutions, we refer the reader to \cite{yang2022sok}.

\subsection{Proposer-Builder Separation and MEV-Boost}
Proposer-Builder Separation (PBS) is a novel design concept introduced for Ethereum to address the potential risk of validators becoming excessively sophisticated in block construction \cite{pbs, pbs2}. This approach introduces a clear division of tasks within the block production process, assigning separate roles to proposers and builders. The core responsibility of the proposer revolves around consensus participation, while the intricate task of block construction is outsourced to specialised builders. By integrating PBS into Ethereum's core protocol, validators can function solely as consensus nodes and source the execution payload from the builder's market on the execution layer. This enables validators to profit from MEV available in the mempool without direct involvement in the complexities of block building. This shifts the competition for MEV and its inherent centralising influences from the validator level, closely related to the consensus layer, to the builder level, primarily associated with the execution layer.

Within the framework of PBS, builders are involved in a competitive environment where they strive to offer the maximum value to proposers. Their bidding capacity depends on their ability to construct a valuable block using transactions from their exclusive order flow and public mempool. Determining the optimal set of transactions or bundles that yield the highest value necessitates using complex algorithms, which may consume significant energy resources. Builders willingly undergo this computational cost to secure a monopoly over a single slot duration without engaging in staking or, previously, mining activities as consensus participants do.

Yet, the successful integration of PBS into the Ethereum core protocol relies on addressing the complexities surrounding trust assumptions and commitments between proposers and builders \cite{monnot_notes_2022}. As a temporary PBS solution, Flashbots introduced MEV-Boost \cite{flashbotsmev-boost_2023}, which employs relays \cite{relay} as intermediaries between proposers and builders. Relays gather blocks from builders in the form of bids, validate them, and present them to proposers. Consequently, they function as a data availability layer, safeguarding proposers from Denial-of-Service (DoS) attacks stemming from excessive bid submissions.

While PBS aims to counteract the centralising impacts of MEV, MEV-Boost leads to new censorship-related risks, as relays could selectively withhold bids from certain builders to specific proposers. However, in the absence of reliable relays, proposers can turn to their execution clients for local block building. Hence, in the MEV-Boost architecture, relay competition promotes honest behaviour as their activities are subject to public auditability.

\subparagraph{Architecture}
The architecture of MEV-Boost \cite{mevboost} is designed around three key players: validators, who control the proposers, relays, and builders, who prepare the execution payloads. To qualify for receiving execution payloads from builders, validators need to register with their chosen relays and provide details such as the payment address, block gas limit, and their public key \cite{relay}. The block proposal process then unfolds as follows, when a validator is scheduled to propose a block \cite{blockproposal, schwarzschilling2023time}:

\begin{enumerate}
    \item Builders, beginning from the preceding slot, check if the validator for the forthcoming slot is registered with a relay they submit bids to. If so, they start assembling execution payloads using transactions from the public mempool and their private order flow, if it exists.
    \item Builders then forward the prepared execution payload, along with their bid for the validator, to the relays they work with.
    \item Once the relays receive the submitted payloads, they validate them and make them available for the validator.
    \item The validator's MEV-Boost middleware collects headers and associated bids from the registered relays and provides the validator with the header offering the highest value.
    \item The validator blindly signs the header and sends it back to the MEV-Boost middleware, which forwards it to the relevant relay.
    \item The relay then verifies the validator's signature on the signed header and propagates the complete block to the rest of the network.
\end{enumerate}

\subparagraph{MEV-Boost Analytics}
Following the merge in September 2022, MEV-Boost has risen to prominence as the primary solution for validators' block production. Analysis and data presented in \cite{noauthor_mevboostorg_nodate,wahrstatter_time_2023, wahrstatter_mev-boost_nodate} clearly showcase the widespread adoption of MEV-Boost among block proposers. As of November 2022, the proportion of blocks produced by MEV-Boost utilising validators persistently exceeds 85\%, even peaking beyond 90\% by June 2023 \cite{wahrstatter_mev-boost_nodate}. The remaining validators presumably depend on local block construction via their execution clients.

The adoption of MEV-Boost has triggered substantial compensations for validators, surpassing an aggregate of 215k ETH \cite{wahrstatter_mev-boost_nodate}. However, it is important to note that the MEV-Boost ecosystem has exhibited a degree of centralisation. Presently, ten active relays participate in MEV-Boost, with \textit{Ultra Sound Relay}, \textit{Flashbots}, \textit{Agnostic}, and \textit{BloXRoute Max Profit} contributing significantly to the block production, constituting 32\%, 25\%, 17\%, and 10\% respectively \cite{wahrstatter_mev-boost_nodate}. This centralisation around specific relays can be attributed to the positive reputation and trust placed in these relays by both builders and validators.

Additionally, a centralisation trend is evident within the builder market as well. Four leading block builders-\textit{Flashbots}, \textit{Beaverbuild}, \textit{Builder0x69}, and \textit{rsync-build}-cumulatively account for nearly 75\% of all MEV-Boost blocks \cite{wahrstatter_mev-boost_nodate}. This pattern suggests that certain builders may have access to exclusive order flow in addition to the transactions available in the public mempool.

% \section{Consensus}
% \section{What is the reward structure for validators?}
% {\color{red}{we already do this before}}
% It would be interesting to analyze the following rewards:
% \begin{itemize}
%     \item Reward from Staking
%     \item Reward from Gas
%     There may be a relationship between the time delay of a block and the size of the Gas reward. How full are the blocks of ethereum? If we show, that gas reward is small this is not very important
%     \item Reward from MEV
%     Well we do that in the rest of the document
% \end{itemize}
% We do not need an empirical analysis....

\section{Data Collection and Processing}
\label{sec:methodology}
To comprehensively analyse the impact of waiting times on the value accrued by proposers, attestations, and consensus stability, we collected data from the MEV-Boost protocol and the Ethereum consensus layer. Our methodology involved leveraging the public data endpoints presented by the MEV-Boost relays, which furnish information regarding submitted builder bids\footnote{\url{https://flashbots.github.io/relay-specs/\#/Data/getReceivedBids}} and proposed blocks\footnote{\url{https://flashbots.github.io/relay-specs/\#/Data/getDeliveredPayloads}}. Our data extraction process primarily focused on three dominant relays \cite{wahrstatter_mev-boost_nodate}: Ultra Sound, Flashbots\footnote{For Flashbots bids and blocks, we also used the data dumps available at \url{https://flashbots-boost-relay-public.s3.us-east-2.amazonaws.com/index.html}}, and Agnostic. Specifically, we procured bid and proposed block data covering a slot range from 6\,087\,501 to 6\,100\,000, constituting 12,500 slots during the time frame of March 27 to 28, 2023. To fetch the consensus-related data such as attestations, forked blocks, and consensus rewards, we utilised the API endpoints provided by beaconcha.in\footnote{\url{https://beaconcha.in/api/v1/docs/index.html}}. 

We leveraged the builder overview found in mevboost.pics\footnote{\url{https://mevboost.pics}} to identify the builders. Notably, some builders adopt a strategy to submit identical bids (those with matching block hash and value) repetitively until the conclusion of the block auction. Builders are also observed to be submitting the same bid to multiple relays to enhance their selection likelihood. We aimed to pinpoint unique bids in our time-based bid value analysis, which necessitated the removal of duplicates within a single relay and across multiple relays. For each relay, the arrival time of a bid was identified by its first appearance, and we disregarded subsequent duplications from the same builder for the same slot. When aggregating bids across all relays, the earliest bid from any relay was considered to be the first, thereby indicating the initial contribution of a particular bid. 

Across the range of all analysed slots, we discovered an average of 81, 119, and 47 duplicate bids per slot submitted to Agnostic, Ultra Sound, and Flashbots, respectively, resulting in an overall average of 83 duplicates. The sheer volume of submissions-with each relay receiving an average of 647 bids per slot-underscored the necessity for accurate identification and removal of these duplicate bids, a step critical to upholding the accuracy of our results. Figure \ref{fig:dups} located in Appendix~\ref{sec:appdups} provides an example of how the duplicate bids are distributed within a particular slot for each relay.

\section{Results}
\label{sec:results}
Our results, a combination of empirical on-chain data analysis and agent-based model simulations, deeply examine the complex dynamics of waiting games on the Ethereum blockchain. Leveraging datasets from the MEV-Boost protocol and the Ethereum consensus layer, we compiled a robust dataset that encapsulates block production activity over a specific period. Our analysis spotlights essential elements such as the value of waiting, the relative importance of MEV rewards versus consensus protocol proposal rewards, and the relationship between timing, attestation shares, and block orphaning likelihood. 

Moreover, we utilise an agent-based simulation model tailored for PoS-Ethereum to inspect how waiting strategies influence emerging consensus properties. This approach allows us to examine system behaviour and dynamics when a set percentage of validators engage in waiting games, adopting diverse delay durations. Ultimately, we strive to address the central question: can waiting games be played without disrupting consensus?

\subsection{The Value of Waiting}
In the context of PoS-Ethereum, a validator selected to propose a block holds a monopoly for the duration of their allocated slot. While honest validator specs compliance would require proposing a block immediately at the slot's beginning (at 0ms), a positive correlation between value and time may motivate rational proposers to postpone their block proposal. This section delves into the exploration of value of time and scrutinises the significance of MEV rewards in comparison to proposal rewards from consensus. The goal is to assess whether the strategic delay in block proposals provides a substantial advantage.

\subsubsection{The Evolution of Value over Time}

\begin{figure}[htbp]
\centering
\includegraphics[width=\linewidth]{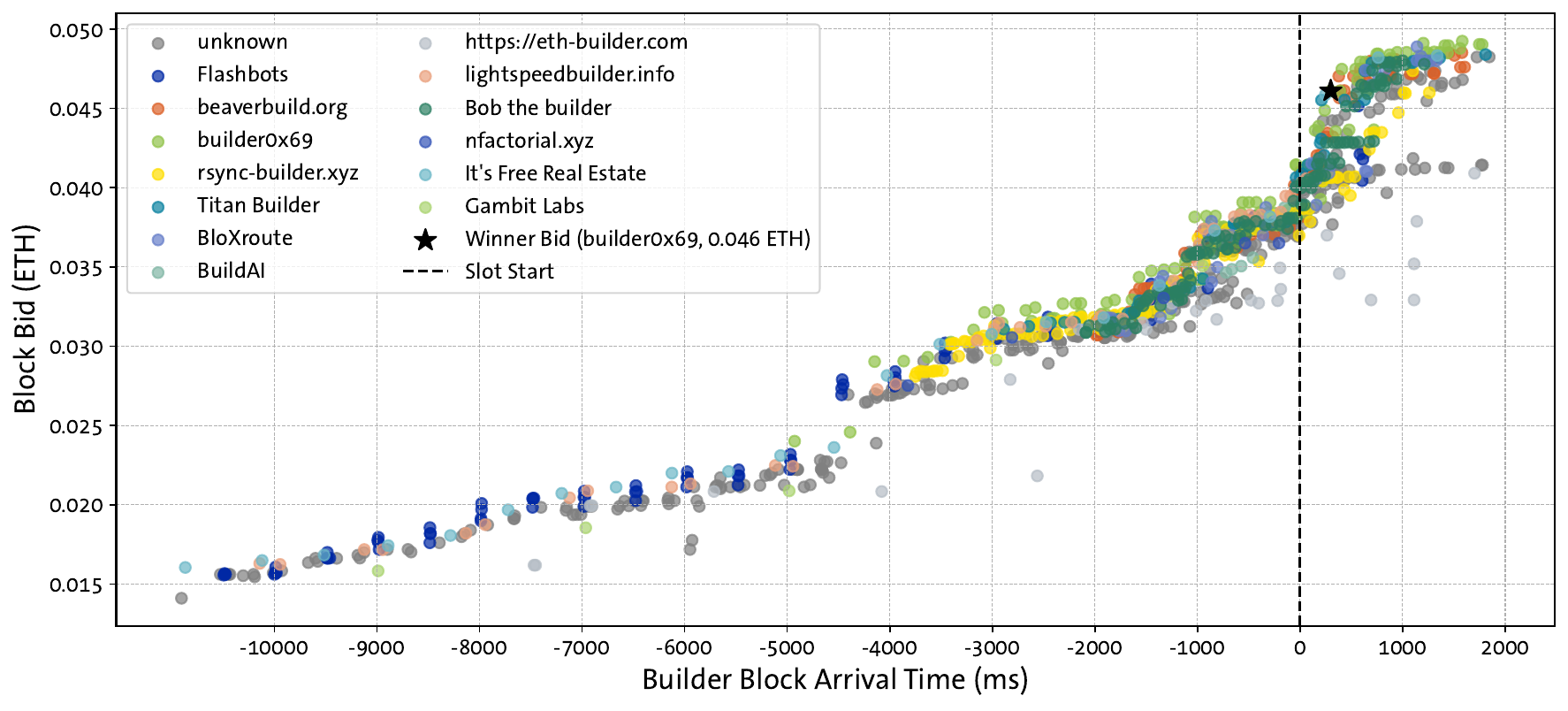}
\caption{The distribution of unique builder blocks, based on their arrival time to the relay and their corresponding bid value, which were submitted by builders for slot 6\,093\,815. The distribution is depicted across Ultra Sound, Flashbots, and Agnostic relays.}
\label{fig:bidTimeVal}
\end{figure}

In order to comprehend the potential value of delay for proposers, we have tracked the bids submitted across the three relays under consideration: Ultra Sound, Flashbots, and Agnostic. Initially, we consider all distinct builder bids submitted across these relays, aggregating them for slot 6\,093\,815. Figure \ref{fig:bidTimeVal} dissects the distribution of bids, which are represented as dots and colour-coded based on the submitting builder. The relation between their arrival time at the relay and their associated value is examined. The study uncovers a distinct upward trend in bid values over time. It is important to note that we measure the arrival time relative to the start time of the slot; therefore, any negative time value indicates that the bid arrived during the previous slot. With this in mind, the earliest recorded bid arrived at -10905ms with a value of 0.014 ETH. On the other hand, the winning bid chosen at 299ms had a much higher value of 0.046 ETH. This represents a substantial increase of approximately 228\% from the initial bid, demonstrating the significant potential advantages of waiting. It is observed that different builders adopt distinct strategies, such as Flashbots builders submitting approximately every 0.5s since the start time of the previous slot, in contrast to rsync-builder.xyz or Bob the Builder, who only commence submissions after a specific point in time.

\begin{figure}[htbp]
\centering
\includegraphics[width=\linewidth]{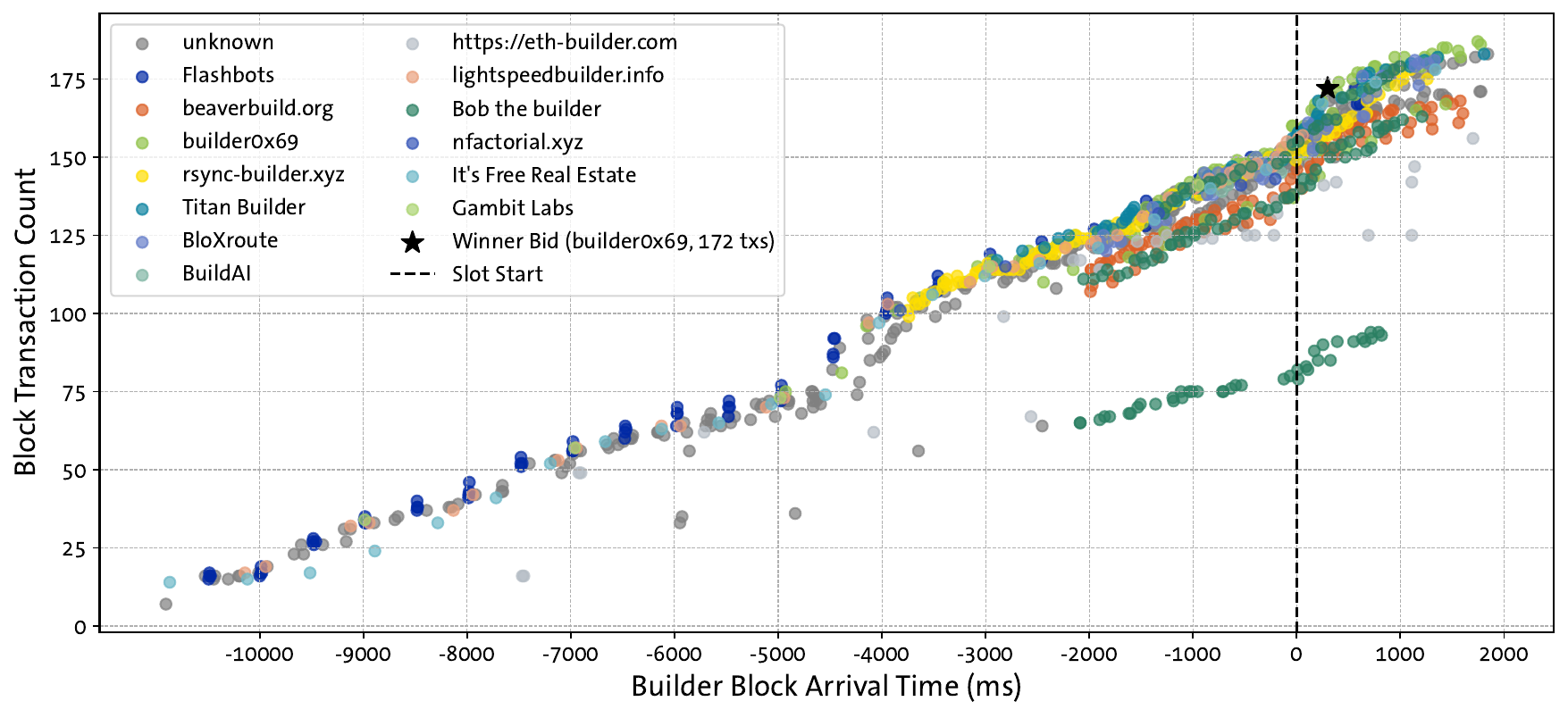}
\caption{The distribution of unique builder blocks, based on their arrival time to the relay and the number of transactions they contain, which were submitted by builders for slot 6\,093\,815. The distribution is depicted across Ultra Sound, Flashbots, and Agnostic relays.}
\label{fig:bidTimeTx}
\end{figure}

This increase in value is attributable to the expanding public transaction pool, and potentially private order flow, that builders observe over time. As new transactions and bundles are successively submitted by regular users and MEV searchers, builders have access to a larger set of opportunities from which to construct their blocks. As a result, they can offer increased value to the validators. The escalating number of transactions included in the builder blocks, as demonstrated in Figure \ref{fig:bidTimeTx}, further substantiates this.

In order to measure the incremental value gained for each millisecond of delay, we collected unique bids from each relay across 750 slots, yielding slightly over 480k unique bids. To accurately capture value progression, we residualized the bid values against slot and builder fixed effects, which might cause artificial value fluctuations due to high or low MEV regimes, as discussed in \cite{schwarzschilling2023time}. Following this, we performed a linear regression on these residualized bid values relative to time, revealing a positive marginal value of $5.71 \times 10^{-6}$ ETH/ms. This supports our initial observation from a single slot, reaffirming that by prolonging their wait time, validators can enhance their MEV payments.

\subsubsection{The Significance of Waiting}
Although we noted a positive marginal value of delay, if the rewards gained from waiting are insignificant compared to the consensus protocol rewards issued for block proposals, then it might not be worth risking the consensus in the first place. To scrutinise the relationship between the rewards from waiting (i.e., MEV rewards) and the proposal rewards from consensus, we examined 5,726 proposed blocks from the relays we have monitored. These blocks belonged to three different epoch ranges between June 1, 2023, and June 11, 2023. While we used the MEV reward data provided by the relays, we fetched the consensus rewards validators received for their proposals from the beaconcha.in API.

\begin{figure}[htbp]
  \centering
  \begin{subfigure}[t]{0.34\textwidth}
    \centering
    \includegraphics[width=\textwidth]{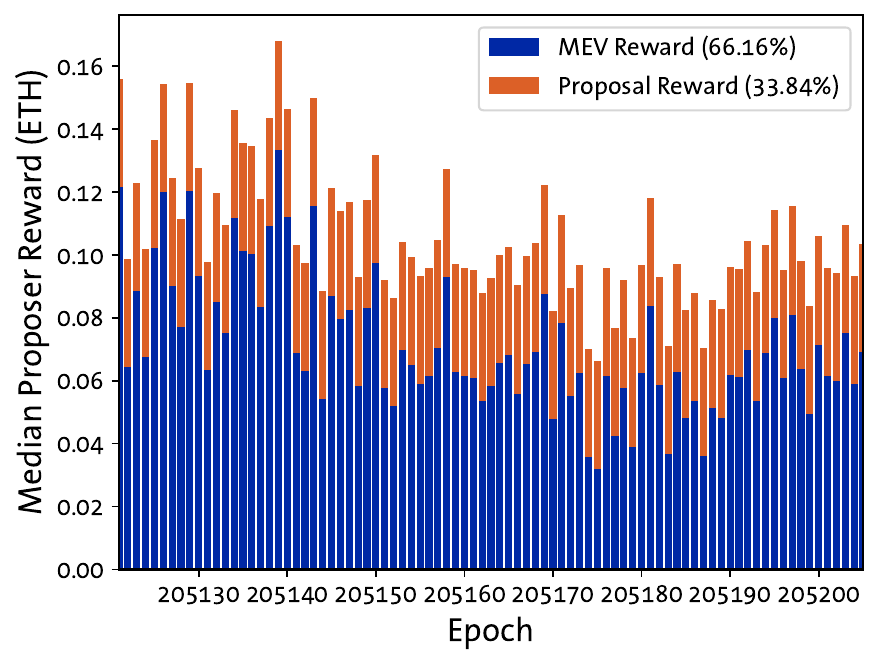}
  \end{subfigure}\hfill
  \begin{subfigure}[t]{0.32\textwidth}
    \centering
    \includegraphics[width=\textwidth]{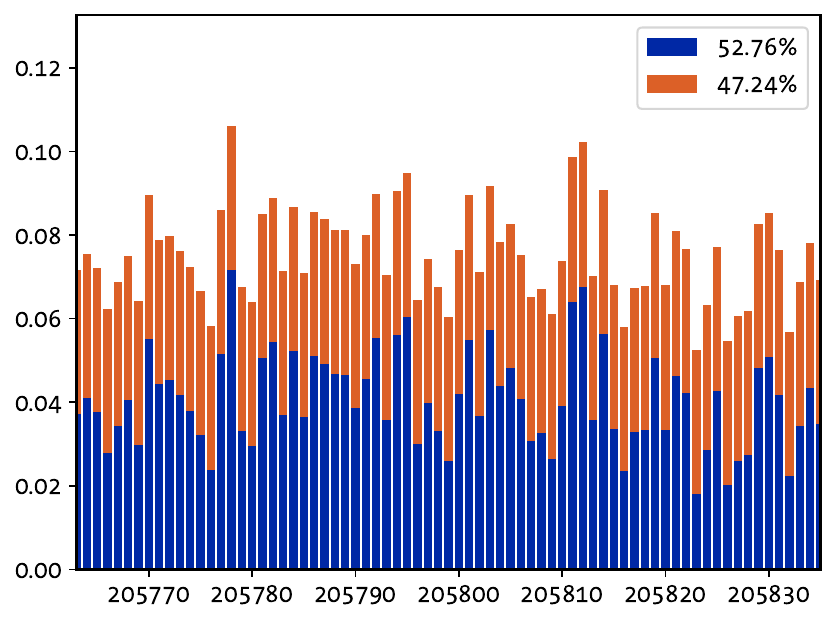}
  \end{subfigure}\hfill
  \begin{subfigure}[t]{0.32\textwidth}
    \centering
    \includegraphics[width=\textwidth]{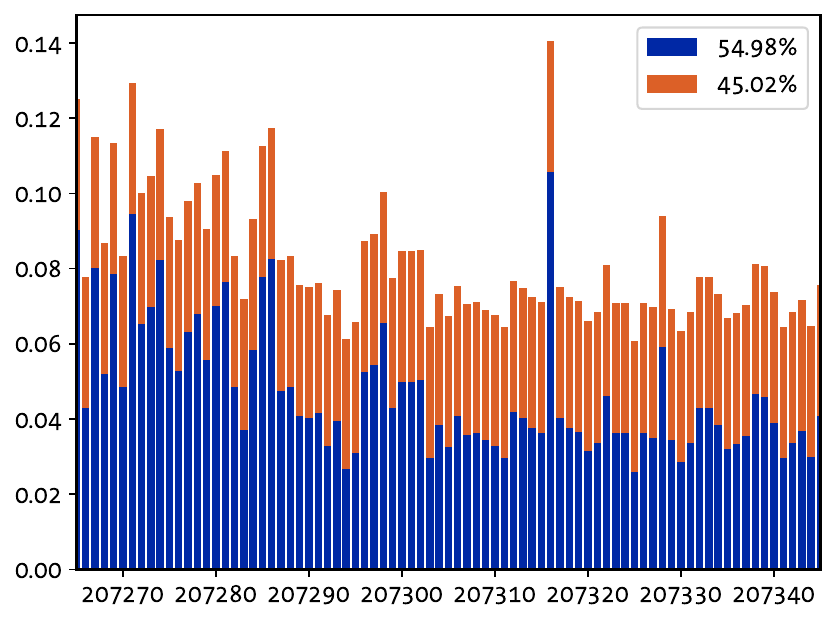}
  \end{subfigure}
  \caption{Median distribution of MEV and proposal rewards from consensus per epoch, visualized across various epoch ranges.}
  \label{fig:rewards}
\end{figure}

In Figure \ref{fig:rewards}, we analyse three distinct epoch ranges, each containin around 80 epochs on average. The diagram distinguishes between two forms of rewards – the blue and orange segments. The blue section represents the median MEV reward that validators, who proposed a block during that epoch via one of our observed relays, received. In contrast, the orange segment represents the median proposal reward issued by the consensus protocol to validators. The proposal reward comprises attestation and sync aggregate inclusion rewards~\cite{cortesgoicoechea2023autopsy}.

A prominent observation across all three epoch ranges is the dominance of MEV rewards over proposal rewards. The first epoch range analysed showcased the largest disparity, amounting to a notable 30\%. Specifically, the median MEV rewards came out to be 0.067 ETH, 0.038 ETH, and 0.042 ETH respectively, culminating in an overall median of 0.048 ETH. Meanwhile, the proposal rewards consistently stayed at 0.034 ETH. This led to a median difference of 0.013 ETH per proposed block. Furthermore, MEV rewards accounted for 58.32\% of all the rewards across the analysed slots.

To determine the median MEV reward since the launch of MEV-Boost (which coincided with the merge), we utilized data from mevboost.pics, revealing 0.053 ETH as the median value. Given that proposal rewards from consensus are anticipated to remain steady at 0.034 ETH, we conclude that MEV rewards have a significant edge over consensus rewards. As a result, the potential gains from waiting outweigh protocol rewards, affirming the value of risking consensus.

\subsubsection{Unrealised Value}
\begin{figure}[htbp]
  \centering
  \includegraphics[width=0.75\textwidth]{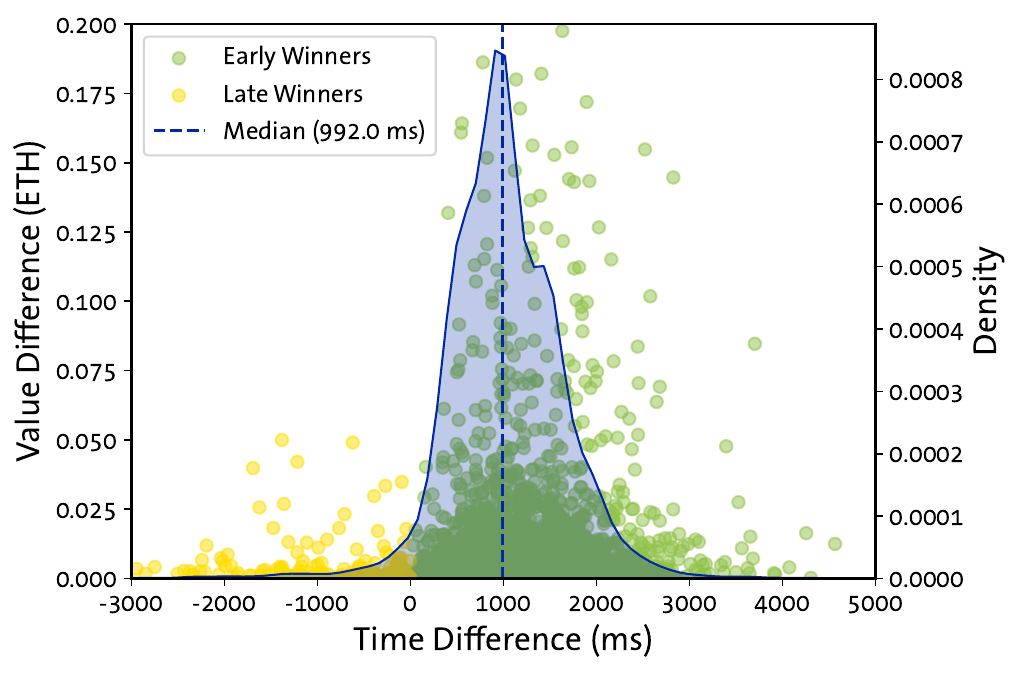}
  \caption{This scatter plot displays early (green) and late (yellow) winning bids in relation to their time and value differences compared to the highest value bid of the same slot. The early bids represent instances where the winning bid reached the relay before the highest bid, while the late ones denote the opposite scenario. The distribution of these early and late bids is supplemented with a density plot capturing the time differential between all bids where the winning bid was not the highest. The plot reveals a median time difference of 992ms, with winning bids arriving before the highest ones. Although the value difference can extend up to 7.61 ETH, this visualisation is limited to a 0-0.2 ETH range for improved readability.}
  \label{fig:unrev}
\end{figure}

Schwarz-Schilling et al.~\cite{schwarzschilling2023time} established that currently, validators do not actively participate in the waiting games, and any delay observed primarily stems from the complex signing processes utilised by certain staking entities and validator clients. Our research supports their findings as we analyse the arrival time of winning builder bids in comparison to the highest value bid observed in that slot. Figure \ref{fig:unrev} portrays the distribution of both early winners (coloured in green), those whose bids arrived prior to the highest bid, and late winners (coloured in yellow), those whose bids arrived after the highest bid. This distribution is presented in relation to time and the value difference between the winning bid and the highest bid. 

Our findings reveal that out of the 8,121 unique winning blocks processed by the relays we have studied, 7,672 (94.47\%) had early winners and 269 (3.31\%) of them had late winners. The remaining 180 blocks (2.22\%) constituted the highest bid itself. The early winners, on median, arrived 1001ms before the highest bid, albeit with a median value of 0.001 ETH less. The late winners, contrarily, arrived 392ms post the highest bid but still delivered 0.0008 ETH less value. In total, across the 7,941 blocks that did not capitalise on the highest bid, with a positive median time difference of 992ms (indicating the winning bid arrived first), a remarkable 931.27 ETH remained unrealised as the highest bid was not available when the winning bid arrived. On the flip side, 27.76 ETH was realisable as the highest bid had already arrived but went unrealised. These results reinforce our contention that validators are not engaging in the waiting game and often act prematurely in selecting the winning bid.

For future work, this analysis should be repeated, incorporating the \textit{getHeader} call timestamp available in the relays. This timestamp signifies the exact moment when the proposer called for the block header (hence the winning bid) from the relay. Comparing this time with the arrival time of the highest value bid could yield more precise figures regarding the total unrealised value and the time difference between winning bids and highest value bids. As this data is not publicly accessible, we resorted to using bid arrival times, which still provide valuable insight.

\subsubsection{Playing the Game Rationally}
Our research thus far has uncovered considerable incentives for rational validators within the Ethereum network to participate in waiting games. However, the manner in which these strategies are employed differs from relay to relay. In the default relay implementation \cite{relay}, each builder bid undergoes a simulation to confirm its validity before it is made accessible to the proposer. This process introduces an average latency of 140ms \cite{optimisticRelays}, shortening the duration of the block auction and reducing the number of competing bids. To counter this latency, the optimistic relay design has been proposed \cite{optimistic-relay-documentation_2023}, and adopted by the Ultra Sound Relay. Under the optimistic approach, it is presumed that builders are submitting valid blocks, which are instantaneously made available while the validation is delayed. However, this strategy necessitates that builders deposit funds upfront, securing payment for the validator if the builder fails to deliver the promised block or payment.

In examining 12,500 slots, we verified the positive influence of optimistic relaying on latency reduction. Out of these slots, 9,250 were relayed using at least one of the three relays we studied\footnote{A builder may have submitted the same winning bid to multiple relays. In such instances, we categorise all relays featuring that bid as relayers of the winning block.}. Table \ref{tab:relay-stats} displays the distribution of block deliveries across the relays. Notably, Ultra Sound Relay, known for its adoption of optimistic relaying, garnered the most bids per slot and delivered the highest quantity of blocks with the largest median value. Additionally, Ultra Sound reported the latest median winning bid arrival time. A more detailed distribution of winning bid timings can be found in Appendix~\ref{sec:relayWinners}.

\begin{table}[htbp]
  \centering
\caption{Summary of Relay Performance Metrics for the Slot Range 6\,087\,501-6\,100\,000.}
  \label{tab:relay-stats}
    \begin{tabular}{lcccc}
    \hline
    Relay & Blocks & Avg. Bids & Median Winner & Median Winner \\
    & Delivered & Received & Arrival Time (ms) &  Value (ETH)\\
    \hline
    Ultra Sound & 3,539 & 549 & 159.0 & 0.0389 \\
    Flashbots & 3,202 & 799 & -450.5 & 0.0348 \\
    Agnostic & 2,509 & 641 & 107.0 & 0.0386 \\
    \hline
  \end{tabular}
\end{table}

These results suggest that by diminishing the block simulation latency, optimistic relaying enables relays to consider more bids and encourages builders to dispatch blocks later in the slot duration. The ultimate consequence is an increasing trend of block auctions being clinched by higher-value, late-submitted bids, thereby augmenting the rewards for validators. While no direct evidence of validators partaking in waiting games has been identified \cite{schwarzschilling2023time}, our analysis of 12,500 slots, along with the extensive historical data provided in \cite{wahrstatter_mev-boost_nodate}, confirms that validators strategically act to maximise their profits from MEV-Boost by registering with relays which deliver the most value. Currently, Ultra Sound Relay leads the pack, delivering around 30\% of all blocks relayed through MEV-Boost, and offering the highest median value of 0.06 ETH to the validators \cite{wahrstatter_mev-boost_nodate}.

\subsection{The Risks of Waiting}
Having established that waiting games present potential profit opportunities for validators, our focus now turns to evaluate the associated risks, particularly in terms of consensus disruption. Although a validator enjoys a monopoly during their slot duration, the ultimate confirmation of the block can hinge on the attestations it gathers from the attestors in the respective slot committees. If a block is proposed too late for attestors to vote on it, these attestors might opt to vote on its parent block instead. Under such circumstances, a block securing less than 40\% of attestations is exposed to the risk of being forked out by the following slot's proposer, using proposer-boost as outlined in Ethereum consensus specifications \cite{ethconsensusspecs}. In this section, we delve into the analysis of attestation shares for the proposed blocks and identify those susceptible to forking. We contrast orphaned blocks with their non-orphaned counterparts to comprehend the potential impact of timing. Moreover, we employ an agent-based model to simulate Ethereum consensus, which allows us to observe consensus stability when a specific portion of validators engage in waiting games, with varying degrees of delay times.

\subsubsection{Attestation Shares}
Expanding on the analysis in \cite{schwarzschilling2023time} about the share of attestations included in the subsequent slot of a proposed block, we examine the next slot attestation shares for the blocks within our data set. We focus on the relationship between the winning bid arrival time to the relay, the attained attestation share, and the vulnerability to being forked out by the next slot proposer through the proposer-boost mechanism \cite{ethconsensusspecs}.

Recalling that the fork choice protocol, LMD-Ghost, computes a block's weight as the cumulative effective balances of validators who attested to the block in a prior slot as an attestation for a block at slot $n$ is only included starting from slot $n+1$, plus the weight coming from the parent block, we assess the impact of the proposer-boost mechanism. This mechanism instantaneously assigns a 40\% committee weight to a timely block proposed within the first 4 seconds of a slot, derived from the total effective balance of validators assigned for that slot, thereby enabling a potential re-organisation (re-org) of a lower-weight past block. To analyse the relationship between this re-org vulnerability and the winning builder block's arrival timing, we measure each block's share of attestations included in the succeeding slot. This involves calculating the block's weight (presuming a uniform effective balance of 32 ETH for each validator) and normalising against the slot's committee weight, which yields the share of attestations a block has accrued from the total potential attestations. Any block with an attestation share below 40\% in the subsequent slot is deemed susceptible to a re-org through the proposer-boost.

\begin{figure}[htbp]
  \centering
  \includegraphics[width=0.75\textwidth]{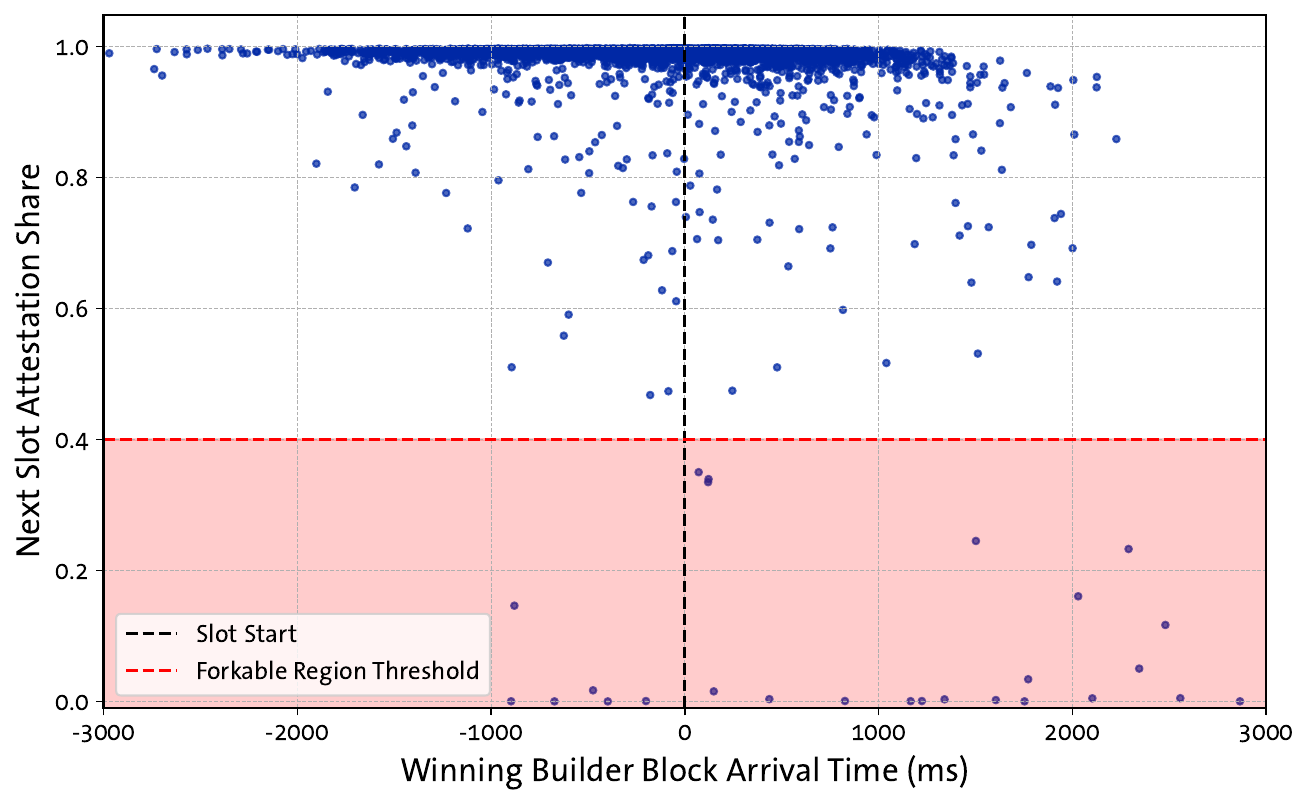}
  \caption{This figure demonstrates the correlation between a block's winning bid arrival time and its attestation share in the subsequent slot. Each blue dot denotes a unique block, with those in the red region indicating blocks vulnerable to getting forked out by the next slot's block proposer due to receiving less than 40\% attestation share.}
  \label{fig:att1}
\end{figure}

Our results highlight that, on average, blocks gather 98\% of attestation shares in the following slot. Out of the 8,121 distinct blocks proposed by the relays under our analysis, only 27 blocks acquired under 40\% shares, with a median arrival time for winning bids of 1223ms. Remarkably, we observed 75 blocks that arrived later than the 1223ms mark yet accrued sufficient shares to be unaffected by the proposer-boost, with an average share of 90\%. This analysis is visually represented in Figure \ref{fig:att1}, where each blue dot stands for a block, and those within the red region identify blocks vulnerable to a re-org using the proposer-boost, based on their attestation share in the subsequent slot.

\begin{figure}[!htbp]
  \centering
  \includegraphics[width=0.75\textwidth]{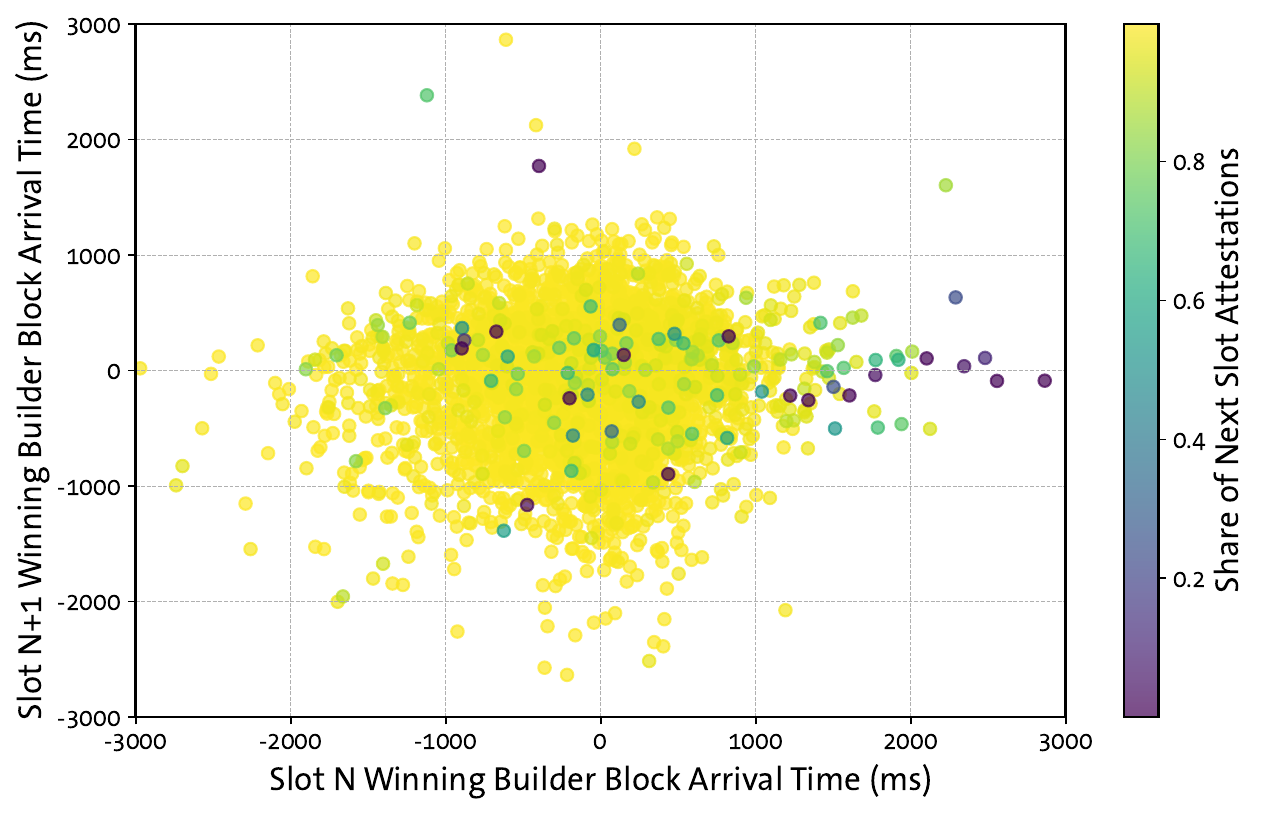}
  \caption{Illustration of the relationship between the arrival times of winning bids in consecutive slots and the subsequent attestation shares. Each dot signifies a block. The x-coordinate denotes the arrival time of the block's own winning bid, and the y-coordinate indicates the arrival time of the following slot's winning bid. The colour gradient reflects the share of attestations a block accrued in the following slot.}
  \label{fig:att2}
\end{figure}

We further examined the relationship between the arrival times of winning bids in consecutive slots and the accrued attestation shares. As depicted in Figure \ref{fig:att2}, each dot corresponds to a block, with its x and y-coordinates representing the arrival times of its own and its following slot's winning bids, respectively. Blocks are colour-coded based on their attestation shares in the following slot. Our results revealed a tendency for attestation shares to decrease when a block's winning bid is delayed while the succeeding slot's winning bid arrives in a timely manner. Specifically, we identified 23 blocks vulnerable to the proposer-boost with attestation shares under 40\% and a following slot's winning bid arriving within the initial 4s. Nevertheless, these blocks remained in the canonical chain, implying that the potential proposer-boost re-org was not exploited.

\subsubsection{Orphaned Blocks}
In our final empirical data analysis, we delved into the occurrence of orphaned blocks within our slot range and its correlation with the timing of the winning bid arrivals. We discovered 151 slots, amounting to roughly 1.2\% of all slots we have analyzed, where a block failed to be included in the canonical chain. Among these, 123 slots had missed proposals, while 28 slots featured orphaned blocks, half of which stemmed from MEV-Boost and the other half were locally built by validators. Figure \ref{fig:forked} in Appendix~\ref{sec:orphaned} illustrates the distribution of these orphaned blocks across relays. 

\begin{figure}[htbp]
  \centering
  \includegraphics[width=0.75\textwidth]{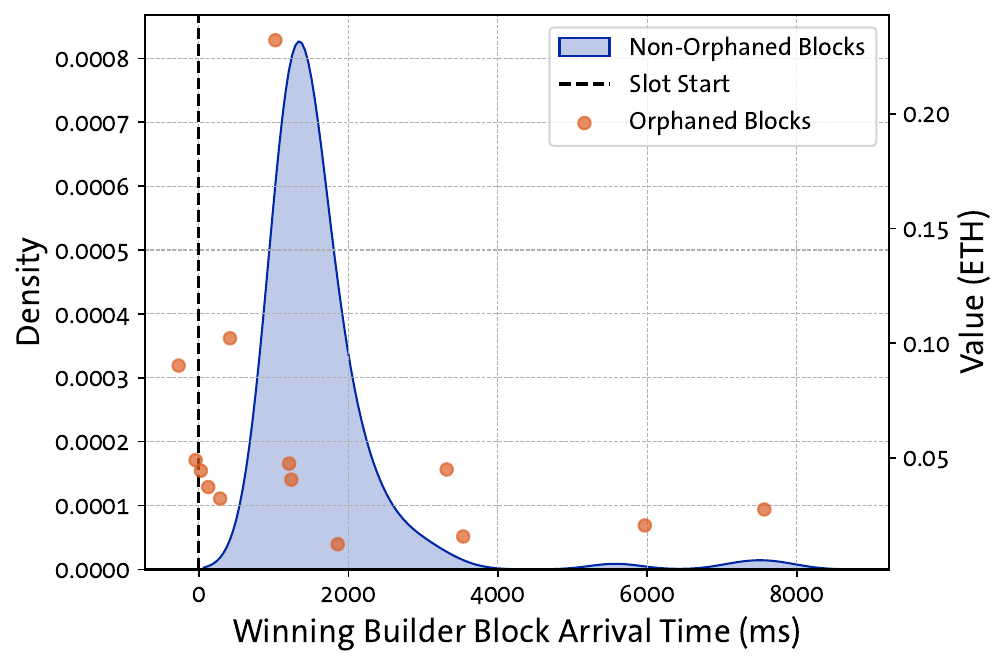}
  \caption{Density plot representing the distribution of non-orphaned blocks that had a winning bid arriving after the earliest orphaned block. Individual orphaned blocks are indicated by scatter points, reflecting their unique arrival times and value.}
  \label{fig:orphaned}
\end{figure}

When we compared the timings of winning bids for both orphaned and non-orphaned blocks, Figure \ref{fig:orphaned} reveal that the winning bid for the earliest orphaned block was submitted 271ms prior to the slot start. We identified a total of 5,631 blocks with winning bids arriving after this point that made it to the canonical chain. Orphaned blocks displayed a median proposal time of 1115ms, contrasting with the non-orphaned blocks' median time of 155ms. Interestingly, 130 of these non-orphaned blocks had winning bids arriving later than the median time of the orphaned blocks.

These observations lead us to two key insights: firstly, orphaning is a relatively rare event, occurring in approximately 0.22\% of the 12,500 slots. Secondly, orphaned blocks are not necessarily late arrivals, as some blocks with later-arriving winning bids avoid orphaning. Thus, we conclude that delaying the selection of the winning bid, and playing the waiting games, does not necessarily result in a block's orphaning.

\subsubsection{The Effects on Consensus: Agent-based Simulation Results}
Until here, we based our observations on the empirical data. In this section, we present an agent-based model to study a scenario where a consistent share of the validators follows the same delay strategy: whenever a delaying validator is selected as a block proposer, they wait until a certain time (which is the same across delayers) into the slot to release the block. They do this to maximise their MEV: from what we observed in the previous sections, there is a good reason to think that MEV rewards correlate positively with the interval between the block release and the previous block release.
For this simulation, we are not interested in estimating the actual amount of increased profit these delayer validators would accrue, as we have already shown that it would be positive. Here we want to observe the eventual fall-out of such a family of strategies on the overall consensus strategy: we want to observe if such strategies may decrease the number of blocks included in the mainchain, which we assume to be an indicator of the consensus efficiency of a blockchain.

The framework we refer to is essentially the same presented in \cite{kraner2023agentbased}: the agents are validators, connected on a peer-to-peer network generated following the Erdos-Renyi random model \cite{erdHos1960evolution}. Time is assumed to be continuous and divided into slots; validators are randomly selected to be block proposers for each slot, and because they are assumed to be fully honest, they release the block exactly at the beginning of the assigned slot. The remaining validators are selected as attestors: they release an attestation to certify on the blockchain that they received the block from the block proposer and that it is valid. If they do not receive the block before the $4$ seconds threshold, they are allowed to attest for the previous head of the chain. The only two random events that may happen are block gossiping and attestation gossiping, which follow exponential waiting times of parameters $\tau_{block}$ and $\tau_{attestation}$. Gossiping events happen when an agent is randomly picked to communicate the information about the blocks/attestations they received to one of their neighbour agents, picked at random as well. In \cite{kraner2023agentbased}, the authors showed how a phase transition in consensus efficiency is observed because of $\tau_{block}$: when the value becomes larger than a certain threshold depending on the topological properties of the peer-to-peer network, the number of blocks included in the mainchain declines abruptly.
In the present work, we increase the number of parameters to take into account the delay strategy as part of the waiting games. We introduce $x^{d}\in[0,1]$ the share of agents who are actively following the delay strategy, and $t^{d}\in[0,12]$ which is the actual delay the delayers wait from the start of the slot before releasing the block.

We proceed by generating a sample of simulations where we vary $x^d$ and $t^d$, while we keep fixed $\tau_{block}=\tau_{attestation}=3$ and the peer-to-peer topology, as well as the simulation time.

As measures we care to observe, we consider the mainchain Rate $\mu$, from \cite{kraner2023agentbased}, to estimate the consensus efficiency, formally defined as:
\begin{equation}\label{eq:mainchain_rate}
      \mu = \frac{|M|}{|B|}
 \end{equation}
 where $|M|$ is the number of blocks included in the canonical mainchain($M$) while $|B|$ is the total number of blocks produced during the simulation.
 
The second measure we consider is the orphan rate of the blocks produced by the delayers $\Theta^{d}$, which we use to proxy, on the one hand, the rewards missed by the delayers and, on the other hand, the direct damage to a consensus as a consequence of the delaying strategy.

\begin{equation}
    \label{eq:orphan_delayer}
    \Theta^{d} = |\{b \notin M \quad s.t.\quad proposer(b) \in \mathcal{D}\}|
\end{equation}
where $\mathcal{D}$ is the set of agents who follow the delayer strategy and $proposer(b)$ represents the agent who proposed block $b$.

\begin{figure}[htb!]
    \centering
    \includegraphics[width=0.45\textwidth]{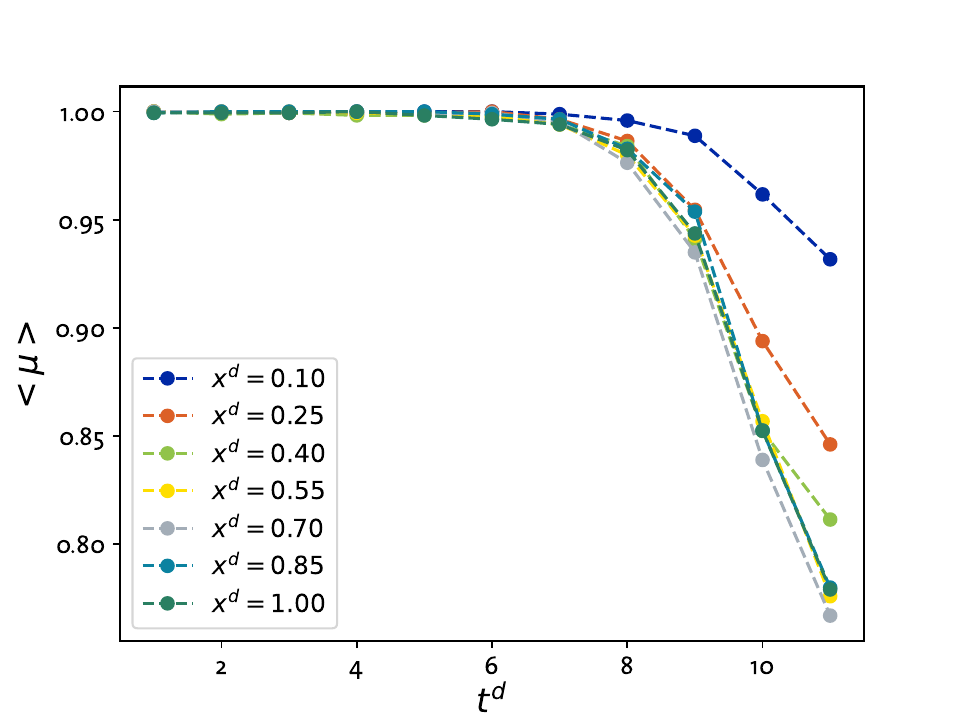}
    \includegraphics[width=0.45\textwidth]{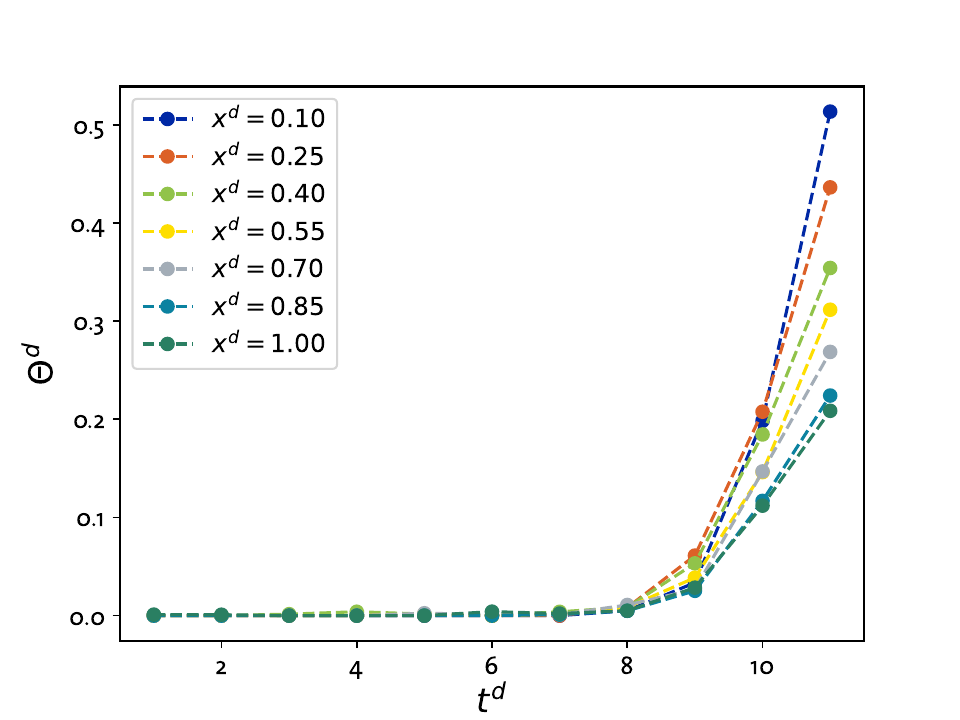}
    \caption{The Mainchain rate (on the left), eq. \ref{eq:mainchain_rate}, and the delayer orphan rate (on the right), eq. \ref{eq:orphan_delayer}, averaged over 20 simulations running for 1000 seconds on an ER\cite{erdHos1960evolution} graph of $N=128$ and $<d>=8$.}
    \label{fig:mev_abm}
\end{figure}

Simulations results are plotted in Figure \ref{fig:mev_abm} and the results are intuitive: the effect of the time delay exercised by the delayers does not significantly affect consensus until it becomes larger than the slot time minus the latency time: $12-3=9$ (the latency also represents the average time between two consecutive gossip interactions between the same two nodes).
We also observe that the effect of the share of delayer stops increasing around $x^{d}=0.5$, at which the number of consecutive mixed blocks (delayer blocks followed by honest blocks) is maximum. This is particularly apparent in the left plot of Figure \ref{fig:mev_abm}, where all share larger than $0.5$ $\mu$ always assume the same value.

While more research is needed to interpret these results, we believe that the results support the hypothesis that a delayer strategy supported by enough validators can be profitable and does not lead to consensus degradation for a range of delay times way more extensive than the ones we observed in the previous sections, in line with the theoretical results on the equilibrium of the waiting games described in \cite{schwarzschilling2023time}.

% \begin{figure}
%     \centering
%     \includegraphics{figures/mev_delayer_orphan_rate.pdf}
%     \caption{The delayer orphan rate, eq. \ref{eq:orphan_delayer}, averaged over 20 simulations running for 1000 seconds on an ER\cite{erdHos1960evolution} graph of $N=128$ and $<d>=8$.}
%     \label{fig:delayer_orphan_rate}
% \end{figure}

\section{Conclusion}
% In this research, we examine the 'waiting games' phenomenon in Ethereum's Proof-of-Stake (PoS) system. We find that validators have substantial incentives, primarily through MEV-Boost, to deviate from the norm and delay their block proposals. Despite this, our findings show only a small percentage of blocks actually included in the Ethereum mainchain contained the highest bid, indicating the unrealized value and the potential benefit of relays like Ultra Sound that use optimistic bid processing.

% Analyzing the associated risks, we ascertain that timing games do not necessarily lead to an increased risk of block orphaning, and even when validators adopt delay strategies, consensus stability is not significantly compromised. 

% In conclusion, our work highlights that the waiting games in Ethereum's PoS mechanism present both potential rewards and manageable risks for validators. Future work will delve deeper into the dynamics among builders, relays, and validators, aiming to optimize waiting games and enhance competitiveness in block auctions.

% NEW

In this study, we have investigated the dynamics of waiting games emerging with Ethereum's transition to a Proof-of-Stake (PoS)-based block producer selection mechanism. We found substantial incentives for validators to deviate from the honest validator specifications, primarily due to the emerging incentive, the Maximal Extractable Value (MEV).

The Proposer-Builder Separation (PBS) implementation by Flashbots, MEV-Boost, have permitted validators to outsource block building to a competitive market of builders. This change has had a significant impact on the dynamics of block proposal strategies. Our results confirm that MEV rewards, which increase over time as more transactions arrive, dominate consensus proposal rewards, which provides sufficient incentive for validators to delay their block proposals - essentially playing the waiting game.

However, timing these delays introduces risk, particularly of block orphaning. Our data demonstrates that while late blocks tend to accrue fewer attestation shares, not all late blocks are at risk of orphaning. In fact, orphaning is a relatively rare event, occurring in only 0.22\% of cases in our study.

We further analyzed the impact of waiting games on consensus stability using an agent-based simulation model, simulating a range of delaying behaviours by validators in various scenarios. Our findings indicate that a delay strategy, if adopted by enough validators, can be profitable without leading to consensus degradation.

In conclusion, our empirical data analysis and simulation results indicate that the risks associated with delaying proposals are outweighed by the potential benefits, given the current dynamics of Ethereum's PoS mechanism and MEV. Looking forward, we aim to further examine the interactions among builders, relays, and validators, particularly in terms of the latencies introduced during these interactions. We hope this will enable a better understanding of how waiting games can be optimized for validators, while ensuring a more competitive block auction for builders.

\label{sec:conclusion}

% \textbf{Acknowledgement} TODO

%%
%% Bibliography
%%

%% Please use bibtex, 

\bibliography{lipics-v2021-sample-article}
\newpage
\appendix

\section{Duplicate Builder Bids by Relays}
\begin{figure}[htbp]
  \centering
  \begin{subfigure}[t]{0.32\textwidth}
    \centering
    \includegraphics[width=\linewidth]{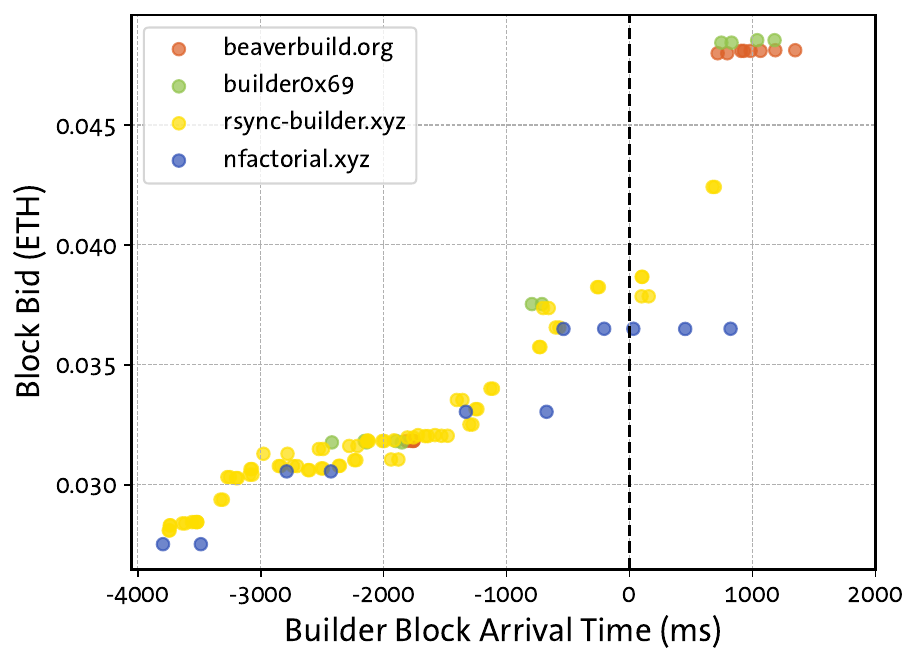}
  \end{subfigure}\hfill
  \begin{subfigure}[t]{0.32\textwidth}
    \centering
    \includegraphics[width=\linewidth]{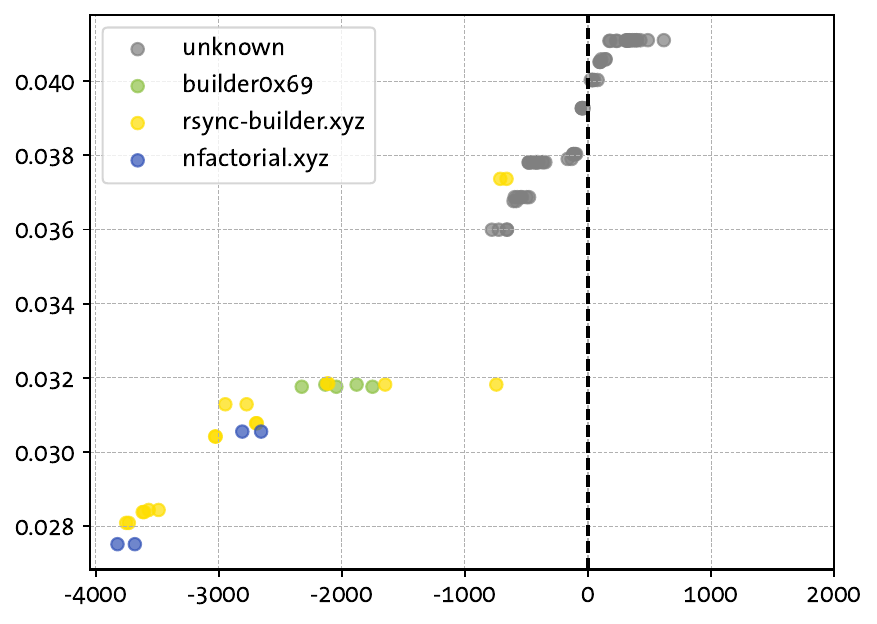}
  \end{subfigure}\hfill
  \begin{subfigure}[t]{0.32\textwidth}
    \centering
    \includegraphics[width=\linewidth]{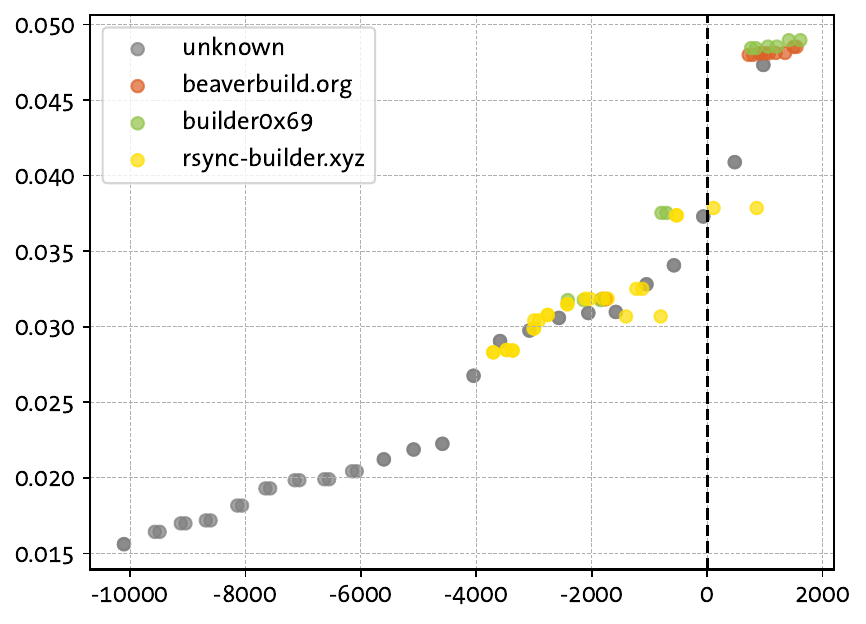}
  \end{subfigure}
  \caption{Duplicate bids per relay for slot 6\,093\,815: Left - Ultra Sound Relay, Center - Flashbots Relay, Right - Agnostic Relay}
  \label{fig:dups}
\end{figure}

Figure \ref{fig:dups} visualizes the distribution of duplicate bid submissions across different relays for a single slot. On an average, a relay is found to receive about 83 duplicate bids. To evaluate the marginal value of time that proposers could potentially gain by waiting, we cleaned the bid data by removing these duplicate entries, allowing us to focus on the contributions of unique bids.

\label{sec:appdups}

\section{Winner Bid Arrival Times by Relays}
\label{sec:relayWinners}
\begin{figure}[htbp]
  \centering
  \includegraphics[width=0.75\textwidth]{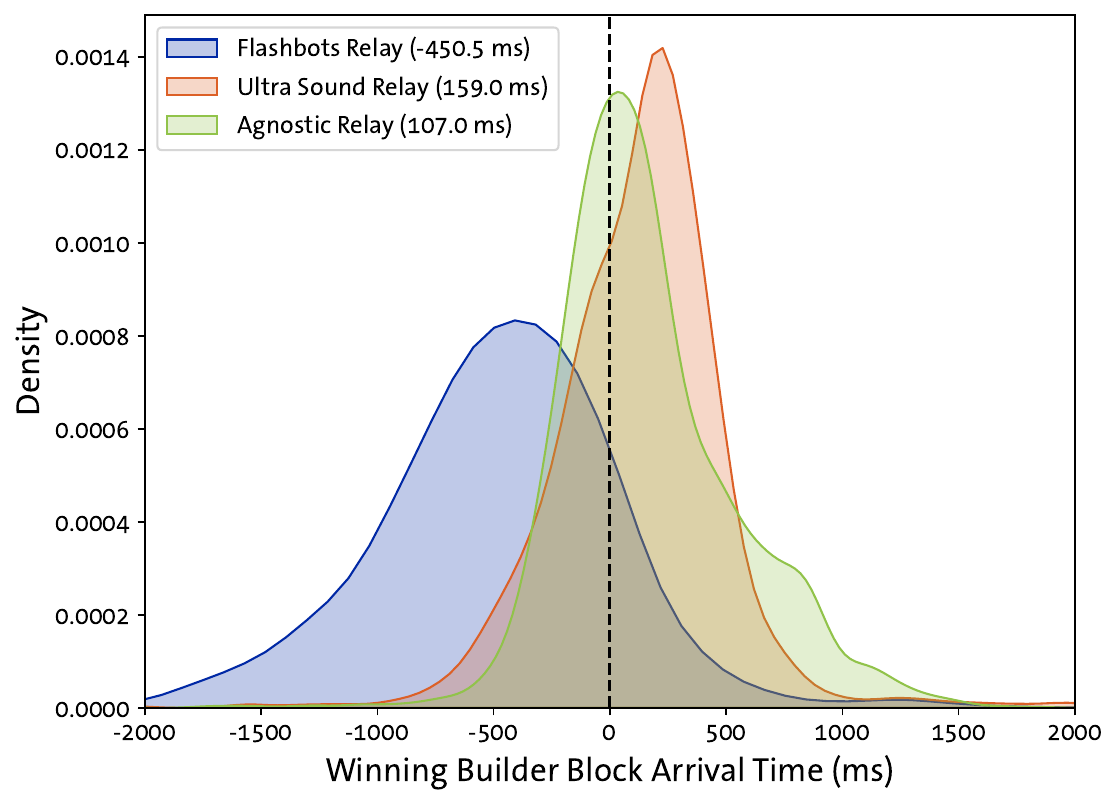}
  \caption{Density plots representing the arrival time distribution of winning bids across different relays. Each plot corresponds to a specific relay, providing a comparative visualization of median arrival times for winning bids.}
  \label{fig:relayWins}
\end{figure}

Figure \ref{fig:relayWins} presents our analysis of winning bid arrival times across different relays, underlining the variations between them. We found that Ultra Sound, recognized for its optimistic nature, and Flashbots, known to be non-optimistic, have distinct median winner arrival times. Interestingly, despite no known claims of being optimistic, Agnostic relay's median winner arrival time is markedly closer to Ultra Sound's than to Flashbots', suggesting potential similarities in their handling of block submissions.

\section{Orphaned Blocks by Relays}
\label{sec:orphaned}

\begin{figure}[htbp]
  \centering
  \includegraphics[width=0.75\textwidth]{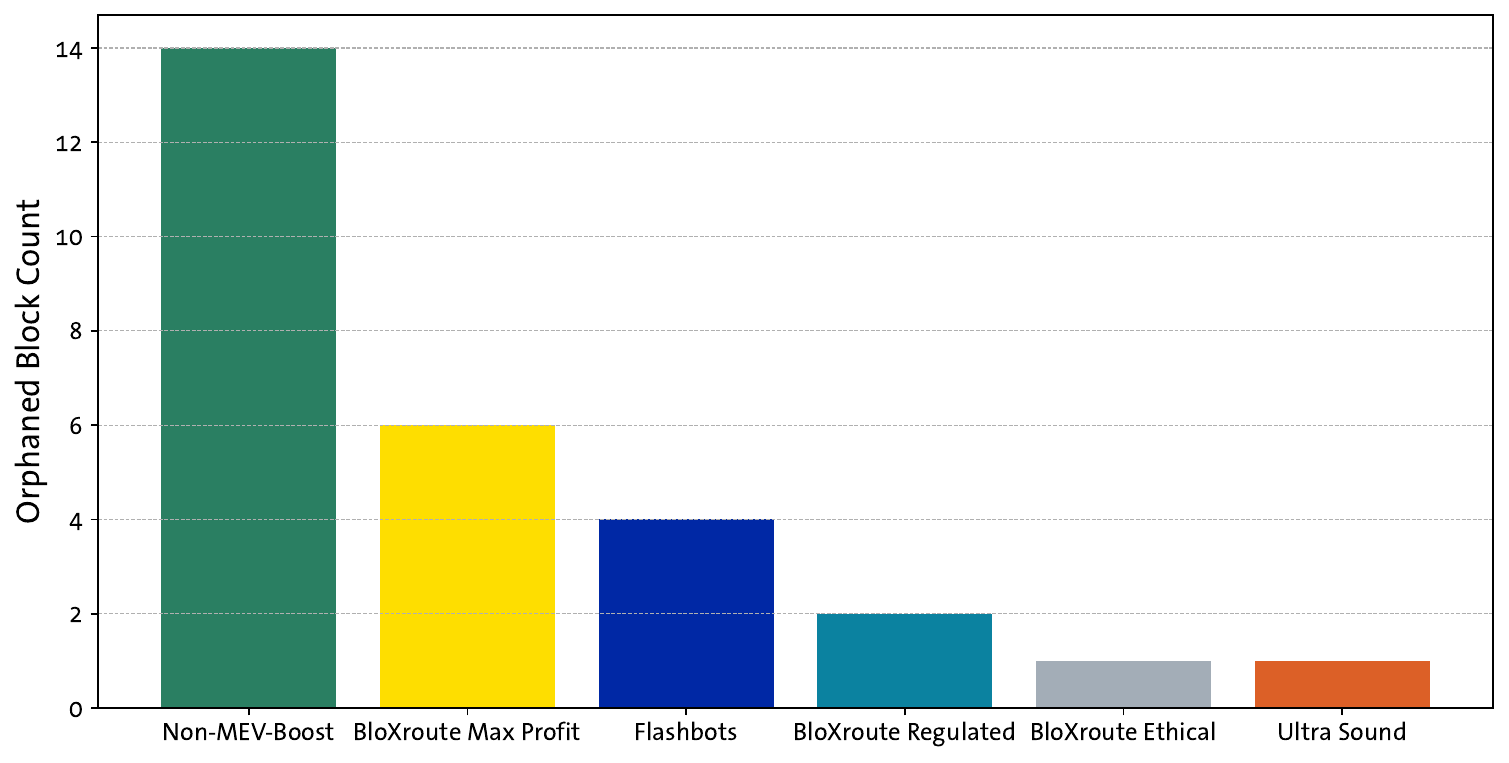}
  \caption{This figure illustrates the distribution of orphaned blocks sourced from non-MEV-Boost using validators and various MEV-Boost relays. Among the 28 identified orphaned blocks, the BloXroute Max Profit relay contributes the most frequently with six instances, with each relay's contribution denoted by the height of its respective bar.}
  \label{fig:forked}
\end{figure}

In the scope of our analysis of 12,500 slots, we identified 28 instances where a block was orphaned, half of which originated from MEV-Boost and the remaining from locally built blocks by validators. The distribution of these orphaned blocks among relays is visualized in Figure \ref{fig:forked}, where BloXroute Max Profit stands out as the relay most frequently associated with an orphaned block, with six occurrences.

\label{sec:forked}

\end{document}